# Towards energy efficient buildings: how ICTs can convert advances?

M. David*,**, A. Aubry*,**, W. Derigent*,**

*Université de Lorraine, CRAN, UMR 7039, Campus Sciences,
BP 70239, 54506 Vandoeuvre-lès-Nancy Cedex
(e-mail: firstname.lastname@univ-lorraine.fr)
**CNRS, CRAN, UMR 7039, France

**Abstract:** This work is a positioning research paper for energy efficient building based on ICT solutions. Through the literature about the solutions for energy control of buildings during operational phase, a 3-layers model is proposed to integrate these solutions: first level consists in communication technologies, second level is about data modelling and third level is related to decision-making tools. For each level, key research topics and remaining problems are identified in order to achieve a concrete step forward.

## 1. CONTEXT AND PROBLEMATICS

Through studies on ICT solutions for energy control of buildings, a 3-layers model is proposed to integrate these solutions and position a new way for energy efficiency.

The building sector is the largest user of energy and $CO^2$ emitter in the EU, estimated at approximately 40% of the total consumption (Sharples *et al.*, 1999). According to the International Panel on Climate Change (European Union, 2010), 30% of energy used in buildings could be reduced with net economic benefits by 2030. Such a reduction, however, is meaningless unless "sustainability" is considered. Because of these factors, healthy, sustainable, and energy efficient buildings have become active topics in international research; there is an urgent need for a new kind of high-technology driven and integrative research that should lead to the massive development of smart buildings and, in the medium term, smart cities. From a building lifecycle perspective, most of the energy (~80%) is consumed during the operational stage of the building (European Union, 2010) (Bilsen *et al.*, 2013). Reducing building energy consumption may be addressed by the physical modifications which can be operated on a building like upgrading windows, heating systems or modifying thermic characteristics by insulating. Another possible path to reduce the energy consumption of a building is to use Information and Communication Technologies (ICT). According to the International Panel on Climate Change, a reduction of energy even greater than the 30% can be targeted by 2030 by considering ICT solutions. In support of this claim, some specialists believe that ICT-based solutions have the potential to enable 50-80% greenhouse gas reduction globally. In this respect, ICT innovation opens prospects for the development of a new range of new services highly available, flexible, safe, easy to integrate, and user friendly (Bilsen *et al.*, 2013).

This, in turn, should foster a sophisticated, reliable and fast communication infrastructure for the connection of various distributed elements (sensors, generators, substations...) that enables to exchange real-time data, information and knowledge needed to improve efficiency (e.g., to monitor and control energy consumption), reliability (e.g., to facilitate maintenance operations), flexibility (e.g., to integrate new rules to meet new consumer expectations), and investment returns, but also to induce a shift in consumer behaviour.

The adoption of ICT within Europe still strongly varies between countries and sectors due to several reasons (Batenburg and Constantiou, 2009). First, user training and awareness in using advanced ICT remains a problematic issue and, consequently, personalized programs adapted to each group of users/citizens still need to be developed (Perera *et al.*, 2013) (Rostocki and Weistroffer, 2014). Second, platform-based ICT often fail to fully consider and involve the Human beings into the decision process, and without appropriate methodologies and services to motivate and support citizens behavioural change, such system will continue to fail. In this regard, it is a very important challenge to further investigate this domain to bring citizens closer to systems and products (including buildings) that they use on a daily basis, and vice-versa (i.e., to bring systems closer to citizens), while fostering the user motivation and confidence in using ICT solutions. Furthermore, at the same time, ICT industries, standardisation bodies and policy-makers are undertaking a series of initiatives to steer the ICT development process with the objective of maximizing its socio-economic value while minimizing the threats related to privacy and confidentiality of data (Miorandi *et al.*, 2012).

According to many studies (Yeh *et al.*, 2009), providing appropriate feedbacks to building occupants can significantly reduce the overall energy consumption (from 5 to 20%). However, only relying on peoples awareness and behaviour may not be an effective approach. Indeed, an experimental study (Rizvi *et al.*, 2012) has shown that more than 30% energy saving was achieved immediately after installing a monitoring system in a residential household, but the percentage reduced to less than 4% one month later.

The section 2 introduces the research roadmap and presents the 3-layers control model. The sections 3 to 5 details current solutions for each one of the 3 presented levels (respectively Network Layer, Data Layer and Service Layer). Finally, the section 6 is dedicated to remaining challenges and discusses on future proposals that should be dealt with to increase energy efficiency in building management.

## 2. RESEARCH ROADMAP

This section first introduces a report produced within European FP7 REEB Project[1] (*European strategic research Roadmap to ICT enabled Energy-Efficiency in Buildings and constructions*) (section 2.1). Then, our original research positioning is presented (section 2.2).

*2.1 Roadmap for Energy-Efficient in Buildings*

The REEB Project, that ended in 2010, provided a report (REEB, 2010) which extensively detailed a new vision of smart building in the short, medium and long term. The intelligent and integrated control of such next-generation systems is one of the priority areas tackled by the report.

As stated by the report, ICT contributions to the energy efficiency of buildings will be possible via a multitude of design tools, automation & control systems, decision support to various stakeholders throughout the whole life of buildings, etc. This topic is in the intersection of three disciplines: building/construction, ICT and energy. Some examples of relevant items for an integrative approach are listed in the Figure 1, below.

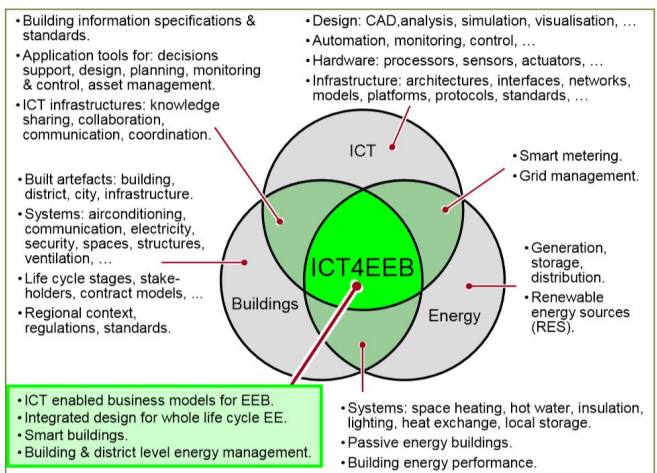

**Figure 1 - Research areas in Energy Efficient Buildings**

For building control, the REEB vision considers that future buildings will be able to communicate and provide information on their status ubiquitously. Real-time information would be available on each component, equipment of the building, and accessible via common protocols for holistic automation & control. Supervision of the building would rely on intelligent systems, able to combine information for all connected devices, from the Internet or from energy service providers. From this aggregated source of information, these intelligent systems would efficiently control HVAC, lightning, and hot water systems along with production, storage and consumption devices inside the building, while into account the user location, needs and wishes. They would then obviously rely on strong data and knowledge management systems, capable to model, gather and aggregate data to produce context-aware information, and on monitoring systems composed of sensors autonomous in energy, that could be embedded in building components. For existing buildings, wireless sensor platforms should be envisaged.

*2.2 An integrated approach based on a 3-layer Model*

Because very few works explore jointly all these aspects (or all the priority areas) of this problem, we try to gather different specialities around a common model.

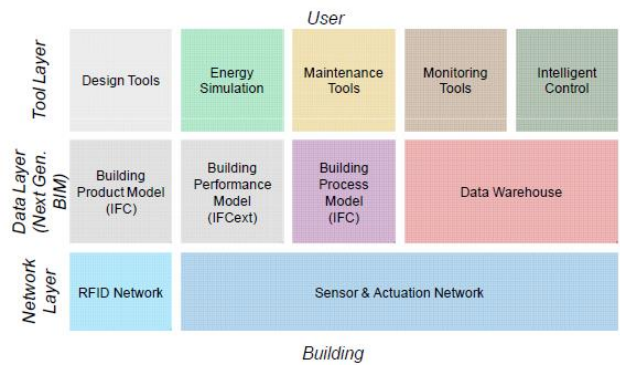

**Figure 2 - 3-layers model between building and user**

Inspired by ITOBO (Information and communication Technology for sustainable and Optimised Building Operation) building management Framework (Guinard et al., 2009) from Cork Institute (see Figure 2), a simplified model ó focusing only on energy efficiency ó can be proposed as depicted on the Figure 3.

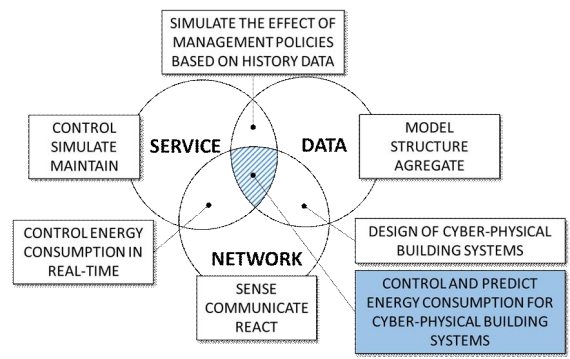

**Figure 3 ó proposed simplified model and targeted area**

This model is composed of three circles, each one representing one layer of the ITOBO model, with dedicated functions. The different layers are overlapping which may lead to additional functionalities. For example, combining a virtual building model (in the õdataö layer) with real-time information coming from the real building (in the õnetworkö layer) can help to design cyber-physical building systems. From this figure, we argue that the objectives targeted by the REEB Project vision can only be fulfilled by functionalities developed in intersection with these 3 circles. This would lead to Cyber-Physical Building Systems (e.g building systems equipped with physical systems ...), controlled energy-efficient intelligent algorithms, capable to predict and simulate the building behaviour.

---
[1] http://cordis.europa.eu/project/rcn/86724_en.html

The aim of the next 3 sections is to make a quick review of existing works. Section 6 provides a synthesis and underlines some research issues.

## 3. NETWORK LAYER

The evolution of communication technologies brings a new way to manage the use of energy in buildings. First, we describe what a "smart building" is (section 3.1), then we synthetize what are the major network's advances in or out of the building (sections 3.2 and 3.3).

*3.1 What is a Smart Building?*

According to the kind of building energy needs, potential reductions are obviously different but a set of new ICT allows to bring the systems more and more "intelligent". The word "Smart" is also used for the Buildings. In 1999, (Sharples *et al.*, 1999) already proposed a definition for the Intelligent Building: "*An Intelligent Building is one that utilises computer technology to autonomously govern the building environment so as to optimise user comfort, energy-consumption, safety and monitoring-functions*". Today this definition seems still valid even if the "autonomous" word became ambiguous regarding the energy optimization. Indeed, the energy efficiency of a building is not any more an isolated problem because the ICT (Internet, Smart Grid) allows to have new outer elements (real-time pricing, weather data integrating, social networks inciting ...) for an autonomous decision-making. 2 generations of Intelligent Buildings can be distinguished. First generation consists of numerous independent self-regulating (automatic) sub-systems. Second generation brings the connection between sub-systems with networking communication.

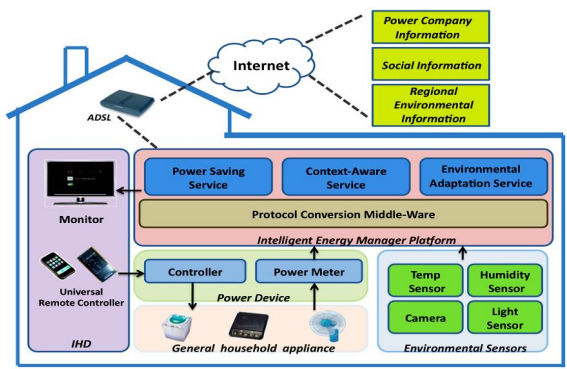

**Figure 4 - Smart Home Architecture** (Kailas *et al.*, 2012)

*3.2 Technologies in the building*

The taxonomy of communication standards for buildings described in (Sharples *et al.*, 1999) is no more relevant because a lot of new technologies emerged especially wireless technologies. (Cheng and Kunz, 2009), or a more recent survey (Kailas *et al.*, 2012) give an overview of the communications and networking technologies. Taken from this last reference, Figure 4 depicted the architecture and Figure 5 gives the main network technologies for communicating in building.

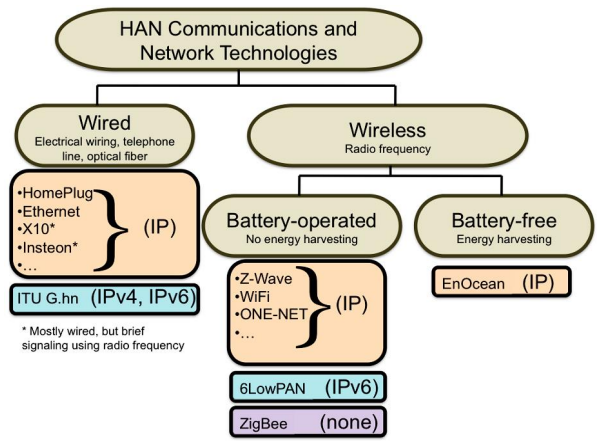

**Figure 5 - Communications and network technologies**

A lot of works have been published in the two past decades about monitoring energy building with Wireless Sensor Network (Calis *et al.*, 2011), (Grindvoll *et al.*, 2012) (Dong *et al.*, 2010). These systems can sense physical environment (temperature, air quality, lightning, occupancy ...). With their processing unit, WSN nodes can also aggregate and compute data, and so offer the possibility to primary react to environmental changes without a central decision (section 5). Because almost 90% of the research works in WSN concern energy control, we can think that this technology is well dedicated to participate to the energy efficiency in buildings.

Because this is a huge market, a lot of companies have tried to develop their own solution (box, software, operating system, or whatever ...). Some examples can be given: Web companies like Amazon (with "Echo") or Apple (with "Homekit") propose a box, which is used to control the overall domestic aspects. First Google's try – PowerMeter (2009-2011) – was a software project to help consumers track their home electricity usage. In 2014, Google acquire Nest Labs and actually exploit a new box called "Home" since the beginning of 2016. Cisco with its Energywise program tried to link network devices and energy control. For some reasons, this system is not well adopted (security or confidentiality problem, incompatibility with the current practices ...). With HomeOS solution (Dixon et al., 2012), Microsoft provides an operating system to enable smarter home for everyone ...

*3.3 Technologies out of the building*

The promising domain of the Internet of Things (IoT) allows to envisage a new way for users to control building automation and also its energy consumption. Many countries have also developed smart metering programs. Smart metering enables more accurate measurement of consumption via the use of advanced meters which are connected to a central unit through a communications network, improving data collection for billing purposes. Smart metering provides information on consumption patterns contributing to more sustainable consumption and energy savings in a Smart Grid context (Gao *et al.*, 2012). Their benefit from the producer perspective is clear but not always for the consumer perspective.

Some advances on game theory usage and/or social networks challenging between consumers can be also a concrete step forward to reduce energy consumption in buildings. The success of "Pokemon" mobile application where there's no concrete thing to win, allows imagining an immersive game where the objective is to progressively reduce the building's consumption. Web technology and social networks addiction has to be used in a long-term strategy to avoid the rough fall of efficiency (30% to 4% of energy reduction after one month) observed in (Rizvi *et al.*, 2012).

Section 3.2 shows a fragmentation of industrial solutions and a real dispersion of adoption efficiency. Section 3.3 reveals not still completely mature technologies. We can hope this is a transitive period, but there is a concrete need for standardisation for all these technologies. Other key points or gaps can be observed in the final section.

## 4. DATA LAYER

The objective of the data layer is to support exchange and management of energy-related data. Solutions to realise these functions are presented in the next subsections.

### 4.1 Exchanging information in BEMS

Current Building Energy Management Systems (BEMS) already communicate with building elements equipped of controllers, such as lights, HVAC, windows or doors. KNX, BACnet or LonWorks (Merz *et al.*, 2009) are examples of existing building protocols, enabling interoperability between intelligent building items. Some, like LonWorks, allow to discover automatically devices added to the building network and gateways are available to pass from a protocol to another. However, the exchanges seem to be limited to simple order ("close the window") or request ("give me the temperature of the room"). Some other protocols coming from the Internet of Things, like O-MI (The OpenGroup, 2014), enables to exchange higher-level requests and demands (e.g "record the temperature between two different dates"). Indeed, communication solutions are already available and efficient. However, most of the time, the control strategy of actual BEMS is often based on very simple rules, fixed and not reactive to the building context. Despite the important quantity of data coming from the building elements, the data management layer itself is often undeveloped.

### 4.2 BIM in the use phase

When addressing the problem of building performance management, one key issue is to define a sufficiently precise thermal model of the considered building. Indeed, it is required to determine the thermal response of the building under different conditions (state of HVAC, positions of windows, value of external/internal temperature ...). Two classic types of methods are used to structure this model (Neto and Fiorelli, 2008). The first one is to identify the building behaviour via AI methods (Neural Networks for example). The obtained models are lacking of adaptability and changing one building parameter (e.g. its user) can totally invalidate them. The second method is to construct a theoretical model of the building that fits the building behaviour. This approach is nevertheless time-consuming for a complete building.

In parallel, the Building Information Modelling (BIM) has been rapidly growing during recent years, and is often used in construction projects during the design phase. BIM is defined as methods and tools aiming at creating building models. These models can be data sets (1D), drawings (2D), CAD models (3D) or building thermal simulation models (4D). These models are precise and scalable. It has been proven on several major BIM projects like the Digital Campus Innovation (DCI) (Shi *et al.*, 2015), a development project involving the use of BIM on a portion of Carleton University's 45 interconnected buildings. However, all these knowledge is often restricted to the design phase, and is not reused during the other building lifecycle phases.

Thus, some works believe it is then interesting to couple these models with real data coming from the smart building elements, to define a "virtual behaviour" of the complete building, via sensors intelligently placed in the building. (Stack *et al.*, 2009) proposes a building performance framework depicted in the Figure 6, where the initial building information model, defined by the design phase, is linked with wireless sensor network and simulation tools to obtain a long-term multi-dimensional simulated and measure data.

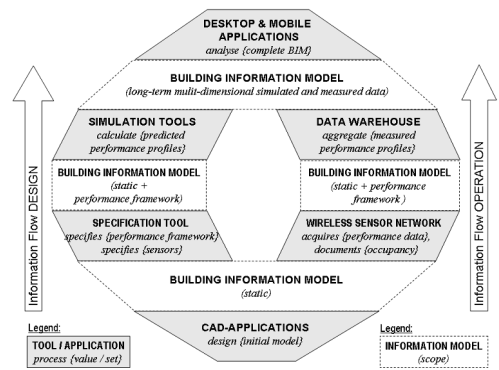

**Figure 6 - Building Performance Framework**

In (Wu *et al.*, 2015), an approach based on a cyber-physical system (CPS) integrating BIM and WSN is proposed, to control the energy consumption of a data centre. This system is then used to run simulations aiming to predict real-time thermal performance of server working environment. The authors argue that such system can be used to quickly pinpoint thermal hot zones and facilitate smarter power consumption. In (Dong *et al.*, 2015), The BIM I² (for BIM enabled Information Infrastructure) is designed for building energy Fault Detection and Diagnosis (FDD) in buildings equipped with HVAC. This infrastructure is composed of 4 interconnected modules (static information, dynamic information, simulation and FDD modules). Static information, describing the building structure with its inner building elements (e.g. Chillers, Air Handler Units or VAV Boxes) with their own properties, is extracted from the BIM models. Dynamic Information is obtained via real-time monitoring performed by building's sensors. Simulation is realised via inputs given by both previous modules, and predicts the nominal building behaviour. FDD then compares the virtual and normal behaviour, to detect faults.

As far as we know, this last article is clearly the most advanced paper on the BIM-based approach. Such infrastructure provides integration between the three circles of the simplified model proposed in the Figure 3. However, to our knowledge, it is the only work tackling this vision and it does not concern energy control strategy but energy fault diagnosis. Further research works are then needed to investigate this new research area.

## 5. SERVICE LAYER

New ICTs, when implemented in buildings, offer the capacity to have a constant up-to-date state of the building (temperatures, humidity level, consumption, presence or absence of users, states of equipmentí ). This is also a necessity and an opportunity for defining high level services for managing the building in terms of energy. Three high level services can be identified as shown in the Figure 7: prediction service, control service, and maintain service. These services should be integrated in a Building Energy Management System. The goal of this section is to focus on the control service only. Concerning the prediction service, the reader is invited to read the reference (Zhao and Magoulès, 2012) that gives an extensive state-of-the-art on this question. Concerning the maintain service, regarding energy efficiency and buildings, maintenance questions are sparsly studied in the litterature. We can cite (Wang and Xia, 2015) as an example of such works.

### 5.1 A service for controlling energy in buildings

Regarding the control service, controlling the energy in buildings is typically a complex decision problem that need to define decision aid services based on artificial intelligence or operational research techniques. That means to be able to: i) define the decision criteria, ii) identify the problem constraints, iii) implement the right technique for finding the right action to be done.

Concerning the criteria, (Kolotsa *et al.*, 2009) gives an overview of the different indices that can be regarded. These indices are decomposed into the following categories: *Energy Use* (annual electricity use in kWh/m²í ), *Global Environment* (Lifecycle environmental impactí ), *Indoor Environment* (Thermal comfort, Indoor air quality, Visual comfort, Acoustic comfortí ), *Cost* (annual ongoing chargesí ), and *Other* (Securityí ). It must be noticed that most of these criteria are competitive. That means that finding a global solution optimizing all of them simultaneously is impossible: a tradeoff must be done.

### 5.2 Solving methods for controlling energy in buildings

Regarding solving methods, several types of methods can be identified. It is not possible for being exhaustive here regarding the state-of-the-art but the following of this section gives two examples for implementing energy management.

In (Ha *et al.*, 2012), the authors propose a three-layers architecture to deal with energy management in buildings: a local layer, a reactive layer and an anticipative layer. The goal of this architecture is to manage what they call services to the user. Three types of services are defined. The end-user service is a service that provides comfort to the user. The intermediate service manages energy storage. The support service produces electrical power to intermediate and to end-user services. Moreover, the service can be qualified as permanent if its energy consumption cover the complete planning horizon. If not, the service is qualified as temporary. The anticipative layer is responsible for scheduling end-user, intermediate and support services taking into account predicted events and costs in order to avoid as much as possible the use of the reactive layer. Typically, the predicted events and costs should come from the Predict service and the Information Structure service (see Figure 7). The reactive layer manages the adjustments of energy assignment in order to follow up a plan computed by the upper anticipative layer despite the unpredicted events and perturbations (typically coming back from the sensors through the data layer on Figure 7). The local layer is composed of devices together with their existing local control systems generally embedded into appliances by manufacturers. It is responsible for adjusting device controls in order to reach given set points in spite of perturbations. As shown by the authors of (Ha *et al.*, 2012), the anticipative layer problem can be modelled and solved as a multi-criteria mixed integer linear program (MCMILP). The criteria are : the cost, the autonomy and the $CO_2$ rejection. In (Ha *et al.*, 2006), the same authors propose a metaheuristic based on a tabu search (TS) for implementing the anticipative layer.

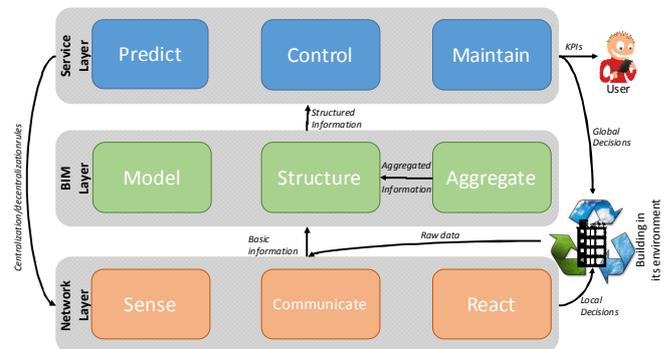

**Figure 7 - 3-layers model information exchanges**

In (Abras *et al.*, 2007), the authors propose a complete multi-agents system for managing the energy in a building. The resource agent calculates the available power resources: to determine what is and what will be the available power. The equipment agent calculates the prediction of power consumption: to determine what are the future power needs taking into account the usual behaviour of users. Then two mechanisms allow to manage energy. The anticipation mechanism is a cooperation between agents for building a consumption plan taking into account the predictions. The emergency mechanism, following a perturbation, update this plan by making cooperate the different agents. The first implementation is completely centralised (MCMILP or TS) whereas the second one is completely distributed (multi-agent system).

# 6. SYNTHESIS AND DISCUSSION

As depicted on figure 7, the 3 layers energy control model is completed with the data exchanges within or between layers. New ICTs for the Network Layer (like WSN) bring a new way for managing energy in buildings. The ability to compute data and to directly react modify the Service Layer decision. Then the future service for controlling energy in buildings will be based on an hybrid approach between centralised and decentralised solving methods.

There are today significant opportunities for efficiency (technological, scientific, economical) and an increase of the user's consciousness (consum'actor). But most are lost due to lack of integration and compatibility. Indeed, energy efficiency in buildings is most of the time a problem approached in a too fragmented way. This paper started a way to reduce the gaps between fragmented parts.

The data layer processes the basic information given by the network layer. Reusing existing BIM models could help to define the virtual thermal behaviors of the building zones (e.g a room, a stair, …). Moreover, it could ease the agregation of building data to generate new knowledge(for example, by grouping the data by building zones and computing agregated values by zone). The connection between models and building controllers, the coherence between building models with different modelling point of views and accuracies, the minimal number of measurement points are all among the scientific questions arising when adressing this challenge.

ACKNOWLEDGEMENT

The authors would like to thank the Agency of Environment and Energy Management (ADEME) for financially assisting this research work, partly funded via the ADEME ITE+ grant.